\documentclass{ws-ijmpd}
\begin{document}
\markboth{P. I. Kuriakose, V. C. Kuriakose} { (Generalized second
law and entropy bound for a Reissner- Nordstr$\ddot{\textbf{o}}$m
black hole)}
\title{Generalized second law and entropy bound for a
Reissner- Nordstr$\ddot{\textbf{o}}$m black hole}
\catchline{}{}{}{}{}
\title{ Generalized second law and entropy bound for a
Reissner- Nordstr$\ddot{\textbf{o}}$m black hole  }
\author{P. I. Kuriakose}

\address{Department of Physics, Cochin
University of Science and Technology, \\
Kochi ,682022, India.\\
sini@cusat.ac.in}

\author{V C Kuriakose}

\address{Department of Physics, Cochin
University of Science and Technology,\\
Kochi 682022, India.\\
vck@cusat.ac.in}
 \maketitle
 \begin{history}
\received{Day Month Year}
\revised{Day Month Year}
\comby{Managing
Editor}
\end{history}

\begin{abstract}

It has been conjectured that black hole interacting with its
surroundings will obey the Generalized Second Law ($GSL$) of
thermodynamics. Conservation of $GSL$ is due to the fully thermal
nature of Hawking radiation and an upper bound on entropy.  We study
these aspects for a Reissner-Nordstr$\ddot{\textbf{o}}$m ($RN$)black
hole and conjecture that $GSL$ may be conserved if the equation of
state of radiation near the horizon is modified. An upper bound on
$S/E$, similar to the Bekenstein form evolves in the calculation.

\end{abstract}
\section{Introduction}
 The studies on black holes during the last 30 years have brought to
light strong hints of a deep and fundamental relationship among
gravitation, thermodynamics and quantum theory. At purely classical
level, the cornerstone of this relationship is the black hole
thermodynamics, where it appears that laws of black hole mechanics
are, in fact, simply the ordinary laws of thermodynamics.
Classically, black holes are perfect absorbers that do not emit
anything; their physical temperature is zero. However, in quantum
theory, black holes emit radiation known as Hawking radiation
($HR$)with a perfect thermal spectrum. This allows a consistent
interpretation of the laws of black hole mechanics as physically
corresponding to the ordinary laws of thermodynamics. There is a
huge increase, of the order of $10^{20}$, in entropy of a star
during its gravitational collapse to become a black hole. It is
presumably associated with the gravitational microstates of the
black hole through the number of ways in which a black hole of a
given mass 'm' and area 'A' can be formed.

The area increase law of black hole mechanics bears a resemblance to
the second law of thermodynamics, that both laws assert that a
certain quantity has the property of ever increasing with time. It
might seem that this resemblance is a very superficial one, since
the area has a geometrical origin whereas second law of
thermodynamics have a statistical origin. Nevertheless, this
resemblance together with the idea that information is irretrievably
lost when a body falls into a black hole led Bekenstein to propose
that a suitable multiple of the area of the event horizon of a black
hole should be interpreted as its entropy and that a generalized
second law ($GSL$) should hold: The sum of the ordinary entropy of
matter outside a black hole plus a suitable multiple of the area of
a black hole never decreases \cite{Be1,Be2}. The $GSL$ directly
links the laws of black hole mechanics to the ordinary laws of
thermodynamics.

It has been proposed that black hole entropy can be due to quantum
entanglement between the interior and exterior states of the black
hole and also that entanglement entropy $S$ is equal to quantum
corrections of Bekenstein - Hawking entropy
$S_{BH}$\cite{Rvb,Lsu,Tmf,Tja}. In these arguments, the black hole
entropy is related to the entanglement entropy in the $QFT$ in the
same spacetime \cite{Sns,Mca}. Even now calculations of black hole
entropy, generalized second law and the conditions to conserve $GSL$
are live subjects in black hole physics.

 The analogy between laws of black hole mechanics and
classical statistical mechanics will break down once the $GSL$ is
violated. Unruh and Wald \cite{Un,Wa} analyzed this situation by
performing a gedankenexperiment. They considered a thought
experiment of lowering a box containing matter toward a black hole
taking into account of the effect of ``acceleration radiation''( the
effective radiation a stationary observer near a black hole would
observe)\cite{Un1}. The resulting change in entropy of the black
hole  $\triangle S_{bh}$ in the round trip process was shown to be
greater than $S_s$, the original entropy of the contents of the box.
The existence of Hawking radiation \cite{Ha} preserves the validity
of the $GSL$ because the thermal radiation is the state of matter
and radiation which maximizes entropy at fixed energy and
volume\cite{Un,Wa}.

In order to make valid the $GSL$,  Bekenstein \cite{Be3,Be4}
proposed a conjecture: There exists a universal upper bound on
entropy $S$ for an arbitrary system of effective radius $R$ and
energy $E$, which can be expressed in Planck units (
$c=\hbar=G=k_B=1 $), as $\frac{S}{E}\leq2\pi R$.

The concept that the thermal pressure of Hawking radiation assumes
infinite value at the horizon is unwarranted, as no physical
pathology is believed to exist at the horizon. To overcome this
problem, many have \cite{Pag,Fro,St,Bi} proposed stress-energy
calculations and obtained a finite value for the temporal and radial
components of stress-energy tensor at the horizon. The information
loss paradox in a black hole can be resolved by treating the Hawking
radiation as not exactly thermal \cite{Haw}, and this concept will
be used in our calculations. This implies that the pressure of
Hawking radiation will have only a finite value at the horizon and
hence the box containing matter can be brought to the horizon. The
state equations of radiation in general are given as
\begin{eqnarray}
\label{1}\rho=\alpha T_r^4; s=\frac43 \alpha T_r^3,
\end{eqnarray}
where, $\rho$ the energy density, $s$ the entropy density, $T_r$ the
temperature of radiation and $\alpha$ a constant. The expression for
the state equations of radiation when applied in the calculations of
gedankenexperiment, it is obtained that the $GSL$ is violated. As
the sanctity of $GSL$ cannot be questioned, the state equations of
radiation ( Eq. (1)) need to be modified.

The knowledge that the Hawking radiation near the horizon is not
fully thermal, leads us to the conjecture that the gravitational
field near the horizon can influence the equation of state of
radiation. The state equations of radiation near the Schwarzschild
black hole were earlier studied \cite{Li}. The motivation for the
present work is to know whether the state equations of radiation for
$RN$ black hole are different from the general expression given in
Eq. (1). The scheme of the paper is as follows. In Sec. 2, we
describe the violation of $GSL$ where ordinary equation of radiation
is used. In Sec. 3, new equations of radiation and the upper bound
are given . In Sec. 4, we give the conclusion.
\section{Violation of Generalized Second Law?}
\subsection{The gedankenexperiment}
A $RN$ black hole with mass $M$ and charge $Q$ is situated inside a
spherical cavity with radius $r_0$, negligible mass and perfect
reflectability. Let us imagine that the black hole and Hawking
radiation be in thermal equilibrium in the cavity. We fill a
rectangular box of volume $a A$ ( $a$ the height and $A$ the cross
section area of the box ) with thermal radiation of temperature
$T_r$ at infinity. Now lower the box adiabatically through a hole on
the cavity to the horizon, release the contents, then slowly raise
the box back to infinity. In general $T_r\gg T_{bh}$. The increase
in the energy of the black hole in the above process is \cite{Un,Wa}
\begin{eqnarray}
\label{2}\varepsilon=E_r-W_{\infty},
\end{eqnarray}
where, $W_{\infty}$ is the work delivered to infinity and $E_r$ is
the rest energy of radiation in the box. We have
\begin{eqnarray}
 \label{3}E_r=\alpha a A T_r^4\nonumber\\
 W_{\infty}=W_1-W_2,
\end{eqnarray}
where, $W_1$ is the work delivered to infinity on account of the
weight of box and radiation and $W_2 $ the work delivered to the
black hole on account of the buoyancy force of Hawking radiation.
The entropy of radiation inside the box is
\begin{eqnarray}
 \label{4}S_r=\frac43 \alpha a A T_r^3.
\end{eqnarray}
Since the process of lowering and raising the box is adiabatic,
$S_r$ remains constant.
\subsection{Calculation of $W_1$}
This is the energy delivered to infinity as the box is dropped on to
the horizon under gravity and may be given as
\begin{eqnarray}
\label{5} W_1=E_r-E;\nonumber\\
E=A\int_l^{l+a}\rho(x) \chi(x) dx.
\end{eqnarray}
On reaching the horizon, the bottom lid of the box is opened so that
the radiation in the box will be in contact with the Hawking
radiation. Then, $E$ is the energy of the radiation inside the box
after it has attained the thermal equilibrium with the Hawking
radiation near the horizon, $l$ is the distance from the bottom of
box to the horizon, $\chi(x)$ is the red shift factor and $x$ is the
proper distance from the horizon to the box. Under thermodynamic
equilibrium between acceleration radiation and radiation inside the
box at a height $l$, the temperature of radiation becomes $T_0(l)$.
Then we have
\begin{eqnarray}
\label{6} \rho(x)=\alpha T_{loc}^4\nonumber\\
\chi(x)=[1-\frac{2M}{r(x)}+\frac{Q^2}{r^2(x)}]^{1/2},
\end{eqnarray}
where $T_{loc}$ is the temperature of acceleration radiation locally
measured. $T_{loc}$ is related to the equilibrium temperature
$T_0(l)$ as \cite{To}
\begin{eqnarray}
\label{7} T_{loc}=\frac{T_0(l)}{\chi(x)}.
\end{eqnarray}
On the horizon ($l=0$), $T_0(l=0)=T_{bh}$. For $l\ll r_H$ and
writing  $r=r_H+x$, $\chi(x)$ may be modified as
\begin{equation}
\label{8}
\begin{array}{c}
\chi(x)=\frac{2^{1/2}(M^2-Q^2)^{1/4}}{r_H}x^{1/2}.
\end{array}
\end{equation}
On substituting Eqs. (6, 8) in the expression for $E$ in Eq. (5), we
get
\begin{equation}
\label{9}
\begin{array}{c}
E=\frac{\alpha A r_H^3}{\sqrt{2l}}\frac{T_0^4}{(M^2-Q^2)^{3/4}}.
\end{array}
\end{equation}
Similarly
\begin{eqnarray}
\label{10} S_r=\frac43 \alpha T_0^3 A
\int_i^{l+a}\frac{dx}{\chi(x)^3}\nonumber\\ =\frac{4\alpha A
r_H^3}{3\sqrt{2l}}\frac{T_0^3}{(M^2-Q^2)^{3/4}}.
\end{eqnarray}
But, in an adiabatic process entropy never changes. So
\begin{equation}
\label{11}
\begin{array}{c}
S_r=\frac43 \alpha a A T_r^3.
\end{array}
\end{equation}
From (10) and (11)
\begin{equation}
\label{12}
\begin{array}{c}
T_0=(2l)^{1/6} a^{1/3}(M^2-Q^2)^{1/4}\frac{T_r}{r_H}.
\end{array}
\end{equation}
In Eq. (9)
\begin{equation}
\label{13} E= (2l)^{1/6}\alpha
a^{4/3}A(M^2-Q^2)^{1/4}\frac{T_r^4}{r_H}.
\end{equation}
But, $E_r=aA\alpha T_r^4$. So
\begin{equation}
\label{14} E=\frac{(2l)^{1/6}a^{1/3}(M^2-Q^2)^{1/4}}{r_H}E_r.
\end{equation}
Now work done to infinity on account of gravity is
\begin{equation}
\label{15}W_1=E_r-E=E_r-\frac{(2l)^{1/6}a^{1/3}(M^2-Q^2)^{1/4}}{r_H}E_r.
\end{equation}

\subsection{Calculation of $W_2$}
The work done on the black hole on account of the buoyancy of
Hawking radiation is \cite{Un,Wa}
\begin{eqnarray}
\label{16}W_2=A \int_l^{l+a} P(x) \chi(x)dx,
\end{eqnarray}
where, $P(x)$ is pressure of Hawking radiation. If the Hawking
radiation is fully thermal, then
\begin{eqnarray}
\label{17}P(x)=\frac13\alpha T_{loc}^4=\frac13\alpha
\frac{T_{bh}^4}{\chi^4(x)}.
\end{eqnarray}
So, at the horizon, $P(x)\rightarrow\infty$. This makes
$W_2\rightarrow\infty $, which means the box cannot be dropped on to
the horizon. In classical gravity, the geometry is treated
classically while matter fields are quantized.

In examining the semiclassical perturbations of the $RN$ metric
caused by the vacuum energy of the quantized scalar fields, we can
treat the background electromagnetic field as a classical field. The
right hand side of the semiclassical Einstein equations will then
contain both classical and quantum stress-energy contributions
\begin{eqnarray}
\label{18}G^{\mu}_{\nu}=8\pi [T^{\mu}_{\nu}+\langle
T^{\mu}_{\nu}\rangle].
\end{eqnarray}
$T^{\mu}_{\nu}$ represents the classical stress-energy tensor of
scalar field and $\langle T^{\mu}_{\nu}\rangle$ is its quantum
counterpart. Now consider the situation where the black hole is in
thermal equilibrium with the quantized field, so that the perturbed
geometry continues to be static and spherically symmetric. To first
order in $\epsilon=\frac{\hbar}{M^2}$ the general form of the
perturbed $RN$ metric may be written as:
\begin{eqnarray}
\label{19}ds^2=-[1+2\epsilon \rho(r)]f dt^2+f^{-1}dr^2+r^2d\Omega^2,
\end{eqnarray}
where, $f=(1-\frac{2m(r)}{r}+\frac{Q^2}{r^2})$ and $[1+2\epsilon
\rho(r)]$ represents the perturbation due to the scalar field. In
order to save the black hole from extinction due to evaporation, the
black hole is assumed to be placed inside a massless reflecting
spherical shell. Inside the shell the quantum field and Hawking
radiation are in thermodynamic equilibrium and hence the black hole
mass function $m(r)$ contains classical mass and the quantum
first-order perturbation. So
\begin{eqnarray}
\label{20}m(r)=M[1+\epsilon \mu(r)].
\end{eqnarray}
This equation explains the back reaction. The metric perturbation
functions, $\rho(r)$ and $\mu(r)$ are determined by solving the
semiclassical Einstein's equation expanded to first order in
$\epsilon$
\begin{eqnarray}
\label{21}\frac{d\mu}{dr}=-\frac{4\pi r^2}{M\epsilon}\langle
T^{t}_{t}\rangle,\nonumber\\
\frac{d\rho}{dr}=\frac{4\pi r}{\epsilon}f^{-1}[\langle
T^{r}_{r}\rangle-\langle T^{t}_{t}\rangle].
\end{eqnarray}
The right hand side of Eq. (21) is divergent on the horizon unless
$[\langle T^{r}_{r}\rangle-\langle T^{t}_{t}\rangle]$ vanishes
there. Not only that both $\langle T^{r}_{r}\rangle$ and $\langle
T^{t}_{t}\rangle$ must be finite at the horizon. The expectation
value of stress-energy tensor of a quantized massive scalar field in
the $RN$ spacetime is given as \cite{An}
\begin{eqnarray}
\label{22}\langle T^{r}_{r}\rangle \mid_{r_h}=\langle
T^{t}_{t}\rangle
\mid_{r_h}\cong\frac{6\pi^3\epsilon}{105m^2(M^2-Q^2)^2}T_{bh}^4,
\end{eqnarray}
where, $m$ the mass of scalar field. Hence, the most probable form
of the Hawking pressure at the horizon is given by Eq. (22)
multiplied by $\alpha$. By using Eq. (16), we now get
\begin{equation}
\label{23}W_2= \frac{4\sqrt{2}\pi^3\epsilon a^{3/2}A}{105 m^2
r_H}\frac{\alpha T_{bh}^4}{(M^2-Q^2)^{7/4}}.
\end{equation}
The increase in the energy of the black hole in the
gedankenexperiment is obtained from Eqs. (2,3,15,23)
\begin{eqnarray}
\label{24}
\varepsilon=E_r-W_{\infty}=E_r-W_1+W_2,\nonumber\\
\varepsilon=\frac{(2l)^{1/6}a^{1/3}(M^2-Q^2)^{1/4}}{r_H}E_r\nonumber\\
+\frac{4\sqrt{2}\pi^3\epsilon a^{3/2}A}{105 m^2 r_H}\frac{\alpha
T_{bh}^4}{(M^2-Q^2)^{7/4}}.
\end{eqnarray}
Near the horizon, $l\simeq0$, hence
\begin{eqnarray}
 \label{25}\varepsilon=\frac{4\sqrt{2}\pi^3\epsilon a^{3/2}A}{105 m^2
r_H}\frac{\alpha T_{bh}^4}{(M^2-Q^2)^{7/4}}.
\end{eqnarray}
The increase of the black hole entropy may be given as,
\begin{eqnarray}
\label{26}\varepsilon/T_{bh}=\triangle
S_{bh}=\frac{4\sqrt{2}\pi^3\epsilon a^{3/2}A}{105 m^2
r_H}\frac{\alpha T_{bh}^3}{(M^2-Q^2)^{7/4}}.
\end{eqnarray}
Since, $T_{bh}\ll T_r$ makes, $\Delta S_{bh}\ll S_r$. This is
violation of $GSL$. In the above calculations, we took, $\rho=\alpha
T_{loc}^4$ and $s=\frac43 \alpha T_{loc}^3$, which are not true,
near the horizon. These equations don't prevail, unless the Hawking
radiation is fully thermal.
 \section{State equations of radiation}
As Hawking radiation is not fully thermal, the box can be brought to
the horizon, where gravity is very strong. This situation, will
change the form of equations of radiation. By the first law of
thermodynamics, we have
\begin{eqnarray}
\label{27}d(\rho V)=T_{loc}ds-p dV.
\end{eqnarray}
This yields \cite{Un}
\begin{eqnarray}
\label{28}\rho+p=s T_{loc},\nonumber\\ dp=s dT_{loc}.
\end{eqnarray}
For a static spacetime, the hydrostatic equilibrium equation,
derived from $\nabla^aT_{ab}=0$, for a perfect-fluid stress-energy
tensor \cite{Un} is
\begin{eqnarray}
\label{29}\nabla_a
p=(\rho+p)[\frac{\zeta^b}{\chi}]\nabla_b[\frac{\zeta_a}{\chi}]\nonumber\\
=-(\rho+p)\frac1{\chi}\nabla_a\chi,
\end{eqnarray}
where, $\zeta^a$ is a static killing vector field. Since the Hawking
radiation satisfies the hydrostatic equilibrium Eq. (29), we have
\begin{eqnarray}
\label{30}\frac{d[\chi(x)p]}{dx}=-\rho(x)\frac{d\chi(x)}{dx}.
\end{eqnarray}
In the flat space situation, the relation connecting $\rho$ and $s$
is given as
\begin{eqnarray}
\label{31}s_r=\frac43\frac1{T_r}\rho_r.
\end{eqnarray}
The term $\frac43\frac1{T_r}$ is the proportionality term connecting
$\rho$ and $s$, which can be expressed as $C(\infty)$. This term is
not a constant, but a parameter depends on the distance from horizon
and may be expressed as $C(l)$, where $l$ is the distance from the
bottom of the box to the horizon. In the spacetime of black hole,
red shift factor also must be taken into account. Therefore, we may
propose that \cite{Li}
\begin{equation}
\label{32}s=C(l) \rho(x) \chi(x).
\end{equation}
This relation will converge to the flat space situation when there
is no gravity. Substituting Eq. (32) in Eq. (28), we get
\begin{eqnarray}
\label{33}\rho(x)+p=C(l)\rho(x)\chi(x)T_{loc}=C(l) \rho(x) T_0;\nonumber \\
p=\rho(x)[C(l)T_0-1].
\end{eqnarray}
From (30) and (33), we get the expressions of radiation in the
context of $RN$ black hole as
\begin{eqnarray}
\label{34}\rho(x)=\rho_0 \chi^{{\frac{-CT_0}{(CT_0-1)}}}\nonumber\\
s(x)=C(l)\rho_0 \chi^{{\frac{-1}{(CT_0-1)}}}.
\end{eqnarray}
Eq. (34) represents the modified state equations of radiation and
are more realistic in explaining the physical situation near the
horizon.  $\rho_0$ is the energy density in the asymptotic limit and
in the asymptotic limit, $\chi(\infty)=1$. Hence
\begin{eqnarray}
\label{35}\rho(\infty)= \rho_0\nonumber\\s(\infty)=
C(\infty)\rho_0=\frac43\frac1{T_r}\rho_0 .
\end{eqnarray}
Eq. (34) converges to flat spacetime equations (Eq. (1)), as
$\chi(\infty)\rightarrow1$. The state equation of radiation in the
context of Schwarzschild black hole had been utilized in calculating
the entropy of self-gravitating radiation systems \cite{Di}. As we
approach the horizon, $\chi\rightarrow0$, hence the energy density
increases but never become infinity because of the thickness of the
box. From Eq. (34) and $S_r=\frac43 \alpha T_r^3 aA$, it can be
shown that
\begin{eqnarray}
\label{36}C(l)\chi^{\frac{-1}{(CT_0-1)}}=\frac43\frac1{T_r}.
\end{eqnarray}
The $R.H.S$ of Eq. (36) is a constant. As $l\rightarrow0$, both
$\chi$ and $\frac1{CT_0-1}\rightarrow0$.  $C(l)$ increases as we
approach the horizon and on the horizon,
$C(l\rightarrow0)=\frac43\frac1{T_{bh}}$.
\subsection{generalized second law}
In calculating the entropy change of black hole, we have earlier
considered the flat spacetime equations of radiation. Now we will
evaluate $W_1$ with the new equation of radiation. We have
\begin{eqnarray}
\label{37}W_1=E_r-E;\nonumber\\E=A\int_l^{l+a}\rho(x)dx\nonumber\\
=A\rho_0\int_l^{l+a} \chi^{-\xi} dx,
\end{eqnarray}
where, $\xi=\frac{CT_0}{(CT_0-1)}$. By substituting Eq. (8) and Eq.
(34) in Eq. (37), we get
\begin{eqnarray}
\label{38}E =
\frac1{(2-\xi)}\frac{2aA\rho_0}{[4(M^2-Q^2)]^{\xi/4}}[r_H/\sqrt{a}]^{\xi}.
\end{eqnarray}
The entropy may be calculated as
\begin{eqnarray}
\label{39}S_r=A\int_l^{l+a}sdx=A\int_l^{l+a}C(l)\rho(x)\chi(x)dx\nonumber\\
=AC(l)\rho_0\int_l^{l+a}\chi^{1-\xi}dx.
\end{eqnarray}
Eq. (39) is evaluated using Eq. (8) near the horizon as ($l\ll r_H$)
\begin{eqnarray}
\label{40}S_r=\frac1{(3-\xi)}\frac{2aAC(l)\rho_0}
{[4(M^2-Q^2)]^{\frac{\xi-1}4}}[r_H/\sqrt{a}]^{\xi-1}.
\end{eqnarray}
But entropy can also be written as, $S_r=\frac43 aA\alpha T_r^3$.
Equating this equation with Eq. (40) and evaluating for $\rho_0$, we
get
\begin{eqnarray}
\label{41}
\rho_0=\frac{(3-\xi)[4(M^2-Q^2)]^{\frac{\xi-1}4}}{{\frac32
C(l)(\frac{r_H}{\sqrt{a}}})^{\xi-1}}\alpha T_r^3.
\end{eqnarray}
In the asymptotic limit, $C(\infty)=\frac43\frac1{T_r}$. We can
calculate the asymptotic value of energy density $\rho_0$ by using
the relation, $\xi(\infty)=\frac{C(\infty)
T_0}{C(\infty)T_0-1}\simeq0$, considering the fact that, $T_r\gg
T_0$. Substituting $C(\infty)$and $\xi(\infty)$ in Eq. (41), we get
\begin{eqnarray}
\label{42}\rho_0=\frac32\frac{r_H/\sqrt{a}}{[4(M^2-Q^2)]^{1/4}}\alpha
T_r^4.
\end{eqnarray}
But $\frac{r_H/\sqrt{a}}{[4(M^2-Q^2)]^{1/4}}$ is a dimensionless
constant and it may be absorbed in $\frac32$. Now substitute
$\rho_0$ in Eq. (38)
\begin{eqnarray}
\label{43}E= \frac32\frac{\alpha
T_r^4}{(2-\xi)}\frac{2aA}{[4(M^2-Q^2)]^{(\xi+1)/4}}[r_H/\sqrt{a}]^{\xi+1}.
\end{eqnarray}
As $l\rightarrow 0$, $\xi(l\rightarrow0)=\frac{ 4/(3T_{bh})T_0}{(4
/(3T_{bh}) T_0-1)}\simeq1$. Energy of radiation near the horizon is
obtained from Eq. (43)as
\begin{eqnarray}
\label{44}E= 3\frac{[r_H/\sqrt{a}]^2}{[4(M^2-Q^2)]^{1/2}} aA\alpha
T_r^4.
\end{eqnarray}
The term $\frac{[r_H/\sqrt{a}]^2}{[4(M^2-Q^2)]^{1/2}}$ is a
dimensionless constant. Had we taken the asymptotic expressions in
calculating the energy of radiation near the horizon, the value
would have been approximately zero. Now in eq. (24)
\begin{eqnarray}
\label{45}\varepsilon =
3\frac{[r_H/\sqrt{a}]^2}{[4(M^2-Q^2)]^{1/2}}aA\alpha
T_r^4+\nonumber\\\frac{4\sqrt{2}\pi^3\epsilon a^{3/2}A}{105 m^2
r_H}\frac{T_{bh}^4}{(M^2-Q^2)^{7/4}}.
\end{eqnarray}
The entropy change of the black hole
\begin{eqnarray}
\label{46}\triangle S_{bh}=
3\frac{[r_H/\sqrt{a}]^2}{[4(M^2-Q^2)]^{1/2}}\frac{aA\alpha
T_r^4}{T_{bh}}+\nonumber\\\frac{4\sqrt{2}\pi^3\epsilon a^{3/2}A}{105
m^2 r_H}\frac{T_{bh}^3}{(M^2-Q^2)^{7/4}}\gg S_r.
\end{eqnarray}
thus conserving the $GSL$.
\subsection{Upper bound on $S/E$}
We have from Eqs. (38, 40)
\begin{eqnarray}
\label{47}\frac{S}{E}=\frac{(2-\xi)}{(3-\xi)}C(l)[4(M^2-Q^2)]^{1/4}\frac{\sqrt{a}}{r_H}.
\end{eqnarray}
RN Black hole temperature is given as
\begin{eqnarray}
\label{48}T_{bh}=\frac{\sqrt{M^2-Q^2}}{2\pi r_H^2}.
\end{eqnarray}
From Eq. (47)
\begin{eqnarray}
\label{49}\frac{3-\xi}{2-\xi}=\frac{E}{S} C(l)T_{bh}\frac{4\pi
\sqrt{a} r_H}{\sqrt{2}(M^2-Q^2)^{1/4}}.
\end{eqnarray}
Near the horizon, $C(l\rightarrow0)=\frac43\frac1{T_{bh}}$. Eq. (49)
is modified with the situation $\frac{3-\xi}{2-\xi}>1$, as
\begin{eqnarray}
\label{50}1<\frac{E}{S}\frac43\frac{4\pi \sqrt{a}
r_H}{\sqrt{2}(M^2-Q^2)^{1/4}}\nonumber\\
\frac{S}{E}<\frac43\frac{4\pi \sqrt{a}
r_H}{\sqrt{2}(M^2-Q^2)^{1/4}}.
\end{eqnarray}
Dimensionally, this formula is of the Bekenstein form
\cite{Be3,Be4}.

\section{Conclusion} Generalized second law must be valid in all
situations. When evaluating the $GSL$, if the asymptotic state
equation of radiation is considered, the $GSL$ will be violated.
Since the Hawking radiation is not fully thermal, the
gedankenexperiment could be conducted close to the horizon, since
the buoyancy force of Hawking radiation is finite at the horizon.
The gravity is so strong near the horizon that the state equations
of radiation must have been affected by it. Here we have obtained
the state equation of radiation near the horizon of a Reissner-
Nordstr$\ddot{\textbf{o}}$m black hole and found that the $GSL$ is
conserved. In the asymptotic limit, the equations converge to the
usual expressions $\alpha T_r^4$ and $\frac43\alpha T_r^3$. The
parameter $C(l)$ connecting the entropy and energy density is
$\frac43\frac1{T_r}$ in the asymptotic limit and
$\frac43\frac1{T_{bh}}$ near the horizon.

In the above calculation, the upper bound on $S/E$ is analogous to
the one given by Bekenstein. The upper bound on $S/E$ is a necessary
condition to have the conservation of $GSL$. The above procedure has
a slight disadvantage that the Eq. (34), doesn't give the exact
value of $\rho$ and $s$ on the horizon because of the coordinate
pathology. It can be obtained only by removing the geometric
singularity at the horizon and will be initiated somewhere else.

\section{acknowledgments}

V.C.K wishes to acknowledge UGC, New Delhi, for financial support
through Major Research project and  Associateship of IUCAA, Pune.

\end{document}